# An Edge-based WiFi Fingerprinting Indoor Localization Using Convolutional Neural Network and Convolutional Auto-Encoder

Amin Kargar-Barzi[1], Ebrahim Farahmand[2], Nooshin Taheri Chatrudi[3], Ali Mahani[2], Muhammad Shafique[4], (Senior Member, IEEE)

[1]Tyndall National Institute, University College Cork, Cork, Ireland
[2]Reliable & Smart Systems Laboratory, Department of Electrical Engineering, Shahid Bahonar University of Kerman, Iran
[3]EMIL lab, College of Health Solutions, Arizona State University, United State
[4]eBrain Lab, Division of Engineering, New York University Abu Dhabi (NYUAD), United Arab Emirates

Corresponding author: Ebrahim Farahmand (e-mail: ebi271@ gmail.com).

This work has been supported in part by the NYUAD Center for Interacting Urban Networks (CITIES), funded by Tamkeen under the NYUAD Research Institute Award CG001, and Center for Artificial Intelligence and Robotics (CAIR), funded by Tamkeen under the NYUAD Research Institute Award CG010. Moreover, the first author acknowledges that this work was conducted while he was a researcher at Reliable & Smart Systems Laboratory, Shahid Bahonar University of Kerman, Iran.

**ABSTRACT** With the ongoing development of Indoor Location-Based Services, the location information of users in indoor environments has been a challenging issue in recent years. Due to the widespread use of WiFi networks, WiFi fingerprinting has become one of the most practical methods of locating mobile users. In addition to localization accuracy, some other critical factors such as latency, and users' privacy should be considered in indoor localization systems. In this study, we propose a light Convolutional Neural Network-based method for edge devices (e.g. smartphones) to overcome the above issues by eliminating the need for a cloud/server in the localization system. The proposed method is evaluated for three different open datasets, i.e., UJIIndoorLoc, Tampere and UTSIndoorLoc, as well as for our collected dataset named SBUK-D to verify its scalability. We also evaluate performance efficiency of our localization method on an Android smartphone to demonstrate its applicability to edge devices. For UJIIndoorLoc dataset, our model obtains approximately 99% building accuracy, over 90% floor accuracy, and 9.5 m positioning mean error with the model size and inference time of 0.5 MB and 51 µs, respectively, which demonstrate high accuracy in range of state of the art works as well as amenability to the resource-constrained edge devices.

## I. INTRODUCTION

Nowadays, users' position related information in indoor environment has received remarkable attention in the majority of applications [1], especially Indoor Location-Based Services (ILBSs). In the contemporary era, ILBSs can be used in various areas such as indoor navigation and tracking, location-based advertising (shopping advertisements), location-based information retrieval (tourists guiding services in a museum, tracking staff and patients in healthcare), and many more [1][2]. An accurate and low-cost localization system is an important component of ILBSs, which has been taken into consideration in academic and industrial sectors. Generally, in the outdoor environment, this issue has been solved by GPS technique, but this method is not a suitable approach for indoor places because of blockage, attenuation or reflection of satellite signals [1]. Consequently, finding an accurate and low-cost indoor localization system is known as an ongoing challenge in this area.

In recent years, different technologies like Camera [3], visible Light [4], Bluetooth [5], WiFi [6], Ultra Wide Band (UWB) [7] and RFID [8] have been used for indoor localization. Among these technologies, WiFi technology has attracted lots of attention because WiFi networks and their infrastructures are available in most public buildings, such as offices and shopping centers. Moreover, most users have a smartphone with Wifi technology. Therefore, the position associated information can be obtained by this technology without any additional hardware and cost.

Traditional localization algorithms, such as trilateration or triangulation, are based on measured information like distance or angle from some references node to estimate the position [9]. These methods need line-of-sight (LOS) communication to measure accurate distances or angles. Hence, it is clear that these methods are not suitable for indoor environments with lots of walls and other types of obstacles [10]. Among different methods, the *WiFi fingerprinting* method can easily overcome the mentioned issue without the need for distance or angle information, so this is a proper method for non-line-of-sight (NLOS) environments. Actually, in WiFi fingerprinting methods, just the Received Signal Strength Indicator (RSSI) of WiFi signals from each Access Point (AP) is used to calculate users' location. In these methods, it is assumed that RSSI of several WiFi signals in one point is unique; therefore, this pattern (i.e. a WiFi fingerprint) can be used to estimate the location [11].

Generally, the WiFi fingerprinting localization has two main phases, including offline and online phases. In the *offline phase*, WiFi fingerprint dataset, also known as radio map, is constructed by collecting RSSI values of accessible APs at several known points in the interested area (each RSSI pattern labeled by its location). In the *online phase*, users utilize the collected dataset to estimate their position. In this phase, a user measures the RSSI pattern at his/her place and sends this data to the system (server or cloud) to find its position by matching the RSSI pattern with the available patterns in the dataset. The matching part aims to find the most similar pattern from the dataset with the measured RSSI pattern.

There are different methods for matching part of WiFi fingerprinting to estimate the position, ranging from probabilistic to K-nearest-neighbor (KNN) and Support Vector Machine (SVM) [12]. These methods require complex filtering and parameter adjustments that are time-consuming and computationally intensive. In order to reduce the time-consuming and intensive computation, Deep Neural Networks (DNN) is recently used in localization [13]–[15]. Although a different number of studies attempt to decrease intensive computation and time-consuming, the main issue still remains to be how to effectively optimized DNN based methods for indoor localization applications. Towards this, in this paper, we propose a light-weight CNN-based model to solve the above issues.

Moreover, most of the state-of-the-art approaches are cloud-based which gather the data and send them to a server/cloud or other devices to analyze and compute the location [16]. *This procedure has the following drawbacks which significantly affect the efficiency of the localization process.*

- Privacy and security concerns by transmitting user's data to third-party platforms, such as a cloud or server.
- Increased latency of the localization process because the data have to be sent to a server for computation followed by the server sending the location response to the user.
- Increased network traffic and system cost coupled with a centralized method that needs a server or cloud for computation.

Recently, edge-based systems are mostly used in different applications in which all computations are performed on edge; so, there is no need to transfer data elsewhere. Therefore, to address the mentioned drawbacks, an edge-based indoor localization system is employed to have a better performance in terms of privacy, latency and cost. However, the edge devices are not able to execute complex algorithms such as complex CNN which were recently proposed for indoor localization. For this purpose, a light CNN-based model is proposed to run on edge devices with limited resources. This suggested model can be implemented on edge devices, such as smartphones, which significantly improves the localization performance in terms of accuracy, latency, and cost. **The overview of our novel contribution is shown in Figure 1 and its main contributions of this article are briefly described in the following.**

- A light CNN-based model to run on the edge device with limited resources in terms of processor, and memory, such as smartphones for indoor localization application. For this purpose, convolutional auto-encoder is utilized for feature extracting, denoising and dimension reduction.
- In terms of pre-processing, the region gridding approach is used to transform the localization from a regression to a classification task and improve network performance by enhancing the localization accuracy and decreasing network size and complexity.
- Evaluate the suggested network model on an android smartphone and validate the scalability of the proposed model by evaluating it on three different public datasets and also our collected dataset named SBUK-D.

**Paper Organization:** The rest of this paper is organized as follows: In Section II, an overview of several related works on indoor localization methods is presented. Then, in Section III, we describe our proposed model and its structure in three main subsections. Afterwards, model evaluation and experimental results are reported in Sections IV and V. Eventually, concluding remarks are drawn in Section VI.

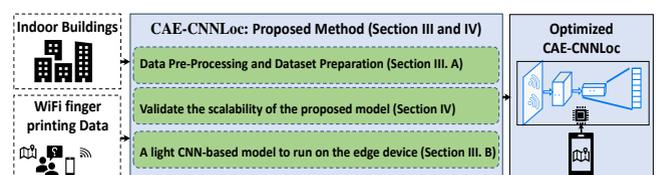

**Figure 1:** The overview of our novel contributions, shown in green boxes

## II. Related work

In the following, some recent indoor localization studies are briefly discussed with their features and problems. Totally, indoor positioning methods are categorized into two main groups named device-free and device-based methods. As is clear from their names in device-free approaches, the location is estimated without any special device carrying by users, for example, using CCTV to detect and locate a person in indoor environments. Conversely, device-based ways need some particular attached devices to compute the location of users [2], for instance, using smartphones WiFi or its other sensors for user localization.

The majority of device-free localization (DFL) techniques need some infrastructures to estimate the position. Some DFL studies, such as [17][18], benefit from wireless sensor network (WSN) for localization which their main idea is to place several wireless sensors around the desired area communicating with each other. These wireless links are affected if a person locates or moves in this area. Hence, the position can be estimated based on these wireless link variations. For instance, authors in [19] used this method and formulated their DFL problem as an image classification problem. Afterwards, a three-layer convolutional auto-encoder neural network suggested extracting features and computing position by raw data with various patterns associated with different positions. Besides, in [20], authors suggested a DFL method by using orthogonal RFID Tags attached on two adjacent walls and utilized some RFID readers to measure phase information and RSSI. The gathered information is fed to a PSO-based algorithm to estimate the position of an object in a 2D place.

It is observed that device-free methods need several extra infrastructures that increase the localization cost. Also, they are not suitable for large buildings, and in most cases, they need LOS communication and can be used for one-floor buildings.

Conversely, device-based localization methods attempt to estimate the position by a device carrying with users. These portable devices utilize various types of modules, such as Bluetooth or Inertial Sensors, to estimate the user location in indoor environments. Nowadays, most people use smart mobile phones with different types of equipment, such as WiFi, Bluetooth and Inertial Sensors; hence, researchers recently attract to benefit from user's smartphones for localization. WiFi-based methods have attracted enormous interest in indoor positioning among various technologies because of their wide availability in most buildings [2]. In WiFi-based methods, some studies like [21] use Time Of Arrival (TOA) technique to estimate the position, while the main disadvantage of TOA is synchronization among all transceivers. In addition, several studies estimate the position based on Angle of Arrival (AOA) approach that requires APs with multiple antennas known as their drawbacks. Besides, some studies locate the user based on WiFi fingerprinting method. The main idea of fingerprinting is to estimate the location by matching the collected RSSI set from surrounding APs named fingerprint with prebuilt WiFi fingerprint dataset [22].

In [23], authors proposed DeepFi for WiFi fingerprinting localization which is a Deep Learning based approach. This method uses Channel State Information (CSI) from all antennas and their all subcarriers which are analyzed with four hidden layers deep network. Generally, CSI-based methods are more accurate than RSS-based ones because they use amplitude and phase of the signal. However, it must be considered that modern smartphones cannot extract CSI; so, it seems that CSI-based methods are not suitable approaches for ILBS.

Unlike CSI-based methods, today's smartphones can easily calculate RSS; hence most prior studies focus on RSS-based WiFi fingerprinting. In this regard, authors in [24] improve the accuracy of WiFi fingerprinting localization by using Weighted K-Nearest Neighbor (WKNN) based on RSSI similarity and spatial position while other KNN-based methods, such as [25], are based on Euclidean distance. In recent years, deep learning methods were used for WiFi fingerprinting. In [23] a deep learning model was suggested for WiFi fingerprinting localization. Moreover, several studies use Auto-encoder (AE) with their DNN model to improve the localization accuracy. For instance, in [26], authors proposed a DNN system for building and floor classification and employ stacked auto-encoders (SAE) to reduce feature space and improve accuracy. Also, a different DNN architecture with SAE was proposed in [27] for multilabel classification of building ID, floor ID and position. Generally, DNN models achieve higher accuracy by increasing their hidden layers. However, a deeper DNN model increases the computational complexity and also computation time. To overcome aforesaid issues, a convolutional neural network (CNN) based structure was proposed in [28] for building and floor classification, which decreases the complexity and reduces the sensitivity of the model to the signal variations. Additionally, CNNLoc method was proposed for WiFi fingerprinting in [13], a CNN-based model including an SAE and one-dimensional CNN. These methods are cloud/server based which all data processing are done on third-party platforms. Hence, the data transmission between user and cloud leads to privacy and security concerns and significant time overhead. A summary of the related works is presented in Table I.

*In summary, the main focus of most studies is to achieve the highest possible accuracy that needs high computation and memory requirements which are mostly based on the cloud/server platform.* So, these cloud/server-based models are not suitable and efficient to directly run on edge devices

Table I
A SUMMARY OF THE RELATED WORKS

| Study | Category | Method | Technique | Attributes and Limitations |
|---|---|---|---|---|
| 15,16,17 | Devised-Free | Wireless Sensor Network (WSN) | Signal variation when an object places between nodes | • Extra infrastructures are needed<br>• Not suitable for large buildings<br>• Need LOS communication<br>• Suitable for one-floor buildings. |
| 18 | Devised-Free | RFID Tags | Phase and RSSI information feed to PSO algorithm to estimate the possition | |
| 19 | Device-Based | Wifi signal | TOA of wifi signal is used to estimate the possition | Need synchronization among all transceivers |
| 21 | Device-Based | CSI-based WiFi fingerprint | Deep Learning (DL) | • CSI-based methods are more accurate than RSS-based<br>• Modern smartphones and devices cannot extract CSI |
| 22 | Device-Based | RSSI-based WiFi fingerprint | Weighted K-Nearest Neighbor (WKNN) | • Modern smartphones and devices support RSS<br>• Not edge based methods a that lead to privacy and security concerns and time overhead |
| 23 | Device-Based | RSSI-based WiFi fingerprint | Euclidean Distance | |
| 13,24,25,26 | Device-Based | RSSI-based WiFi fingerprint | Deep Learning (DL) | |

with limited power, memory and computational resources. In addition, there are some other critical factors for indoor localization, such as cost, latency and privacy/security concerns, which must be considered during system designing. Hence, this study proposes a new edge-based WiFi fingerprinting approach to address these issues along with the highest possible accuracy needed for ILBS in multi-building and multi-floor places.

## III. Proposed method

This section presents a new WiFi fingerprinting system to estimate a person's location in an indoor environment. The proposed architecture is based on CNN classifier that identifies the position of an individual in indoor places. The system diagram of the proposed method is shown in Figure 2. As shown in this figure, the localization system consists of three main phases, including pre-processing, network training and post-training optimization. The details of each part are elaborated in the following.

### A. Data Pre-Processing and Dataset Preparation

The first step is to modify raw input data, such that the data is fed into the proposed network in an appropriate format, and only the relevant data is fed. There are three main phases to prepare data for our network as follow:

1) REGION GRIDING

Generally, users' exact position is not required in the majority of ILBS applications and only the zone or region where the user is located is sufficient [1].

This is a fact that led us to look at the indoor localization problem from a different perspective. And by using this fact we aim to improve the performance of indoor localization in terms of not only accuracy but other important metrics such as latency, privacy and others mentioned earlier.

Therefore, to improve the localization performance, we benefit from gridding technique [22] and divided the location area into some square cells with the length of $L$. More importantly, by region gridding we divide the desire area into several regions thus the localization problem is transformed from a regression problem (exact location) to a classification problem (regions).

This significantly simplifies the localization process meaning that it can be implemented with simpler hardware in comparison to exact localization.

2) DATASET AND INPUT PREPARATION

Generally, WiFi fingerprinting datasets were collected based on the exact location. As mentioned in this study we benefited from region gridding to improve the localization performance. Thus, it is needed to transform the exact location to the region-based form. For this aim, we first divided the dataset area into several zones with the length of L and considered a unique coordinate for each cell representing each point location in the given region.

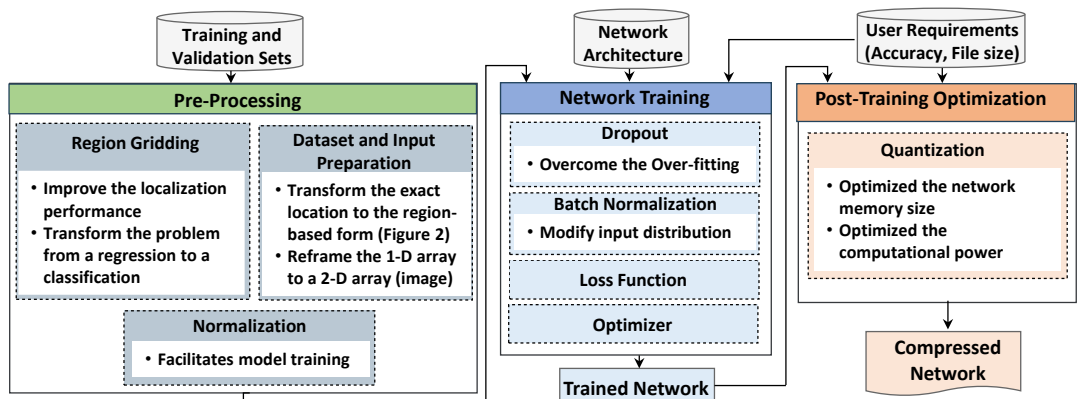

**Figure 2:** Our localization system diagram

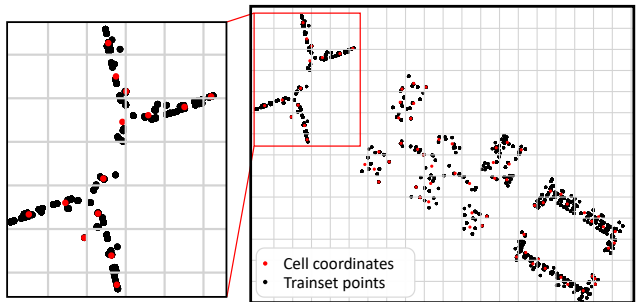

**Figure 3:** Region Gridding of UJIIndoorLoc dataset (train set) by L=20

This procedure for UJIIndoorLoc dataset with *L*=20 is shown in Figure 3. As is evident from this figure, each cell has several samples in which the cell coordinate is assumed for their position. For each cell, this unique coordinate is calculated by averaging all samples' positions inside it. The black points in Figure 3 are the exact location and the red point is the calculated location for each cell. Therefore, all the points in each zone have the same coordination which is the location of the red point.

It must be noted that, in the results section we calculate the distance error between the exact location of each point (black points) and the predicted one which is the location of corresponding cell (red point).

Moreover, in this study, a CNN-based model is suggested for WiFi fingerprinting localization so the model input should be an image. In the WiFi fingerprinting problem, the input is an array of RSSI values, thus we need to reframe this 1-D array to a 2-D array (image) to be compatible with the suggested CNN-based model. Hence, for UJIIndoorLoc dataset, we create a 2-D array from the input, which is a vector with 520 elements. For this purpose, we first add 9 zero elements at the end of each input to have a vector with 529 elements and then reshape the input array to 23x23 2-D array as shown in Figure 4.

### 3) NORMALIZATION

The normalization technique aims at decreasing the input distribution without losing information and facilitates model training. In this paper, input data are Received Signal Strength Indicator (RSSI) value of neighboring access points which are normalized and mapped into [0,1] by the following equation:

$$RSSI_{inew} = \begin{cases} 0, & RSS_i > 0 \ (no\ signal) \\ \dfrac{RSSI_i - RSSI_{min}}{-RSSI_{min}}, & otherwise \end{cases} \quad (1)$$

Where $RSS_{min}$ is the lowest value in the dataset. For example, in UJIIndoorLoc dataset, RSSI values are between $-104\ dbm$ to $0\ dbm$; hence for this dataset $RSS_{min} = -104$.

### B. Network Architecture

In this paper, a Convolutional Neural Network (CNN) is used as the base of the network and a Convolutional Auto-encoder (CAE) is deployed to enhance the performance of the network. In the following, we elaborate the proposed network, which is shown in Figure 6.

### 1) CONVOLUTIONAL NEURAL NETWORK

As mentioned above, this paper proposes a light network for WiFi fingerprinting localization with the highest possible accuracy that can be run on user devices. We leverage a Convolutional Neural Network (CNN) to reduce the input size that is also easier to process without losing important features. Moreover, compared with other machine learning methods, such as SVM and KNN, CNN networks are more robust to the sensitivity of input data variation [28]. This is a critical property in WiFi fingerprinting localization since signal strengths can easily be changed in indoor environments by different factors ranging from multipath effect (as a major factor) or electromagnetic interference to temporary obstacles blocking the WiFi signal [29][30].

As mentioned in sub-section III-A-2, we reshape input data from a vector to a 2D form, so the input can be considered a grayscale image represented radio map in which each pixel is equal to RSSI value from different APs. Hence, the proposed CNN-based network can learn from RSSI values (pixel value) and also the radio maps (pattern) of surrounding APs [28].

### 2) CONVOLUTIONAL AUTO-ENCODER

Convolutional Auto-encoder (CAE) benefits from both CNNs and Auto-encoders (AEs) features. AEs are unsupervised learning methods which are generally used for denoising, dimension reduction and feature extraction by reconstructing the input data in the output.

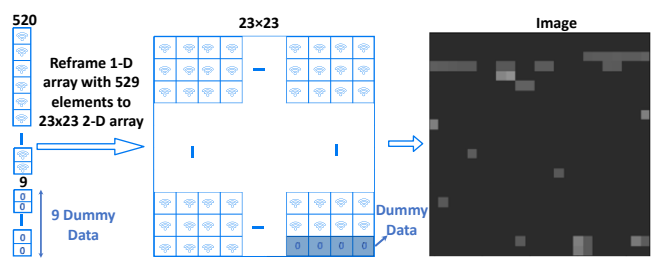

**Figure 4:** Input preparation for UJIIndoorLoc dataset [25]. Reframe 1D array of data included the dummy data to a 2-D array.

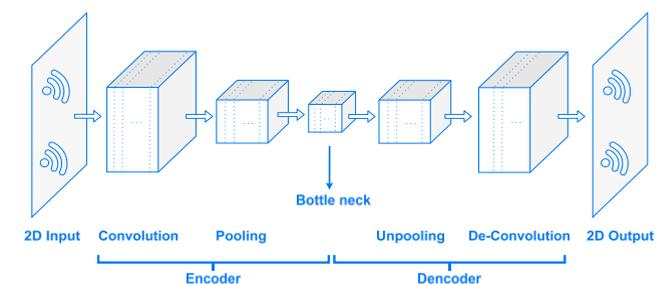

**Figure 5:** Convolutional Auto-encoder

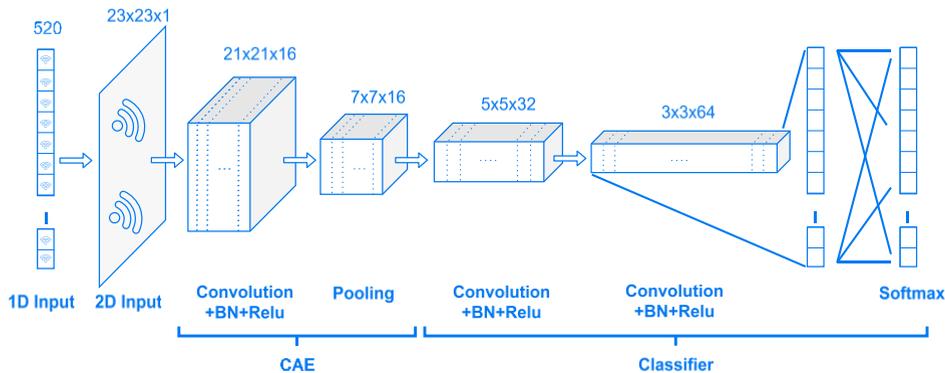

Figure 6: **Proposed network architecture.**

Therefore, in WiFi fingerprinting localization, this is a suitable way to shrink the RSSI value fluctuation caused by different noise sources in indoor environments, such as multipath effect and other sources mentioned before.

Unlike general AE with some fully connected layers, CAE uses some convolution layers which decrease network parameters that subsequently reduce network size. As shown in Figure 5, a CAE has two main parts: a convolutional encoder with some convolution and pooling layers and the complementary deconvolutional decoder with several deconvolutions (transposed convolution) and unpooling (upsampling) layers.

3) **NETWORK DESIGN**

The proposed network architecture is shown in Figure 6. The network comprises Input, CAE and Classifier parts. Input is a grayscale radio map image fed to CAE layer which is indeed the encoding part of CAE that compresses and extracts the main features of the input. Then, the classifier is used to identify the building, floor and location of a user. To enhance the performance of this network we use some layers as follow:

- **Dropout layer:** Dropout is generally used to overcome the over-fitting problem and the main idea is to randomly omit some units in each training iteration; therefore, the network will not train too accurately for the training set which leads to preventing over-fitting. In our proposed network we use this technique before the last layer to avoid over-fitting.

- **Batch Normalization (BN):** This layer is utilized to address the Internal Covariance Shift problem by modification of input distribution in various layers for each mini-batch; hence the convergence rate of the network will be increased by using batch normalization [31]. In the proposed method, BN is used after each convolutional layer, as shown in Figure 6.

C. *Post-training optimization*

In this phase, the proposed model is optimized because it should be implemented on smart mobile phones or other embedded devices with limited memory and computational power. A model with a smaller size not only occupies less storage on the phone but also utilizes less RAM when it runs.

Hence, there is more memory for other applications that improves performance and stability. Besides, a model with lower latency is faster and also has a direct impact on power consumption. It must be noted that generally, post-training optimization decreases model accuracy; thus, there is a trade-off between accuracy and model size or latency which must be considered during the designing process.

For post-training optimization, quantization technique is used [32]. In quantization, the precision of numbers in the model (weights) are reduced to decrease the model size and also computation time. The default type of numbers is float32. In this paper, float16 and int8 quantization are used, and their impact on model performance in terms of model size and latency are evaluated.

**IV. Model Evaluation and Experimental Results**

In this section, the superiority of the proposed CAE-CNNLoc is evaluated in comparison with state-of-the-art methods. To examine the performance of the proposed method, we apply CAE-CNNLoc on UJIIndoorLoc dataset [25]. Moreover, we use the proposed model for indoor localization in our department by collecting its WiFi fingerprinting dataset named SBUK-D as a case study. Additionally, the CAE-CNNLoc network model is implemented with Tensorflow 2.4.0 framework on Google Collaboratory Cloud with Tesla 4 GPU and then its performance is tested on an android smartphone. The network parameters are considered as follows (Table II).

In this study, three various public datasets are used which their details are explained as follows.

**UJIIndoorLoc**: This is the most common dataset for WiFi fingerprinting localization that includes 3 buildings with 4 or 5 floors and covers 108,703 $m^2$ region at the University of

Table II
General network parameters

| Parameter | Value | Parameter | Value |
|---|---|---|---|
| Activation Function | Relu | Optimizer | Nadam |
| CAE Loss Function | MSE | Output Layer Activation Function | Sigmoid |
| CNN Loss Function | Sparse Categorical Cross entropy | | |

Jaume I in Spain. This dataset has training and testing sets with 19,938 and 1,111 samples respectively and each sample has 529 features. The first 520 features show the RSSI value from different APs between -104 dBm to 0 dBm and the null value is shown by 100 that represents inaccessible AP. The location information of each sample consists of Building ID, Floor ID and Longitude - Latitude values in meters [25].

**Tampere:** This dataset was collected at a university building in Tampere, Finland. The building covers approximately 22,750 m$^2$ area with five floors, but just four floors were used to create this dataset. Totally, there are 991 APs in the building; hence in each point, the RSSI value was recorded from these APs. The location in Tampere dataset contains X, Y and Z, which the Z represents the floor [33].

**UTSIndoorLoc:** This dataset was gathered in the FEIT Building at the University of Technology Sydney (UTS). This building covers nearly 44,000 m$^2$ region and includes 18 floors of which 16 floors were used to create UTSIndoorLoc dataset. The position information consists of X, Y and Floor Id, and also each sample has 590 RSSI values from different APs [13].

*A. CAE and CNN optimization*

First, various structures of the proposed model are investigated to find the best possible model for our application. For this aim, different layers such as convolutional and pooling layers with different parameters are tested to achieve the best structure. It must be noted that we only investigate small structures because our goal is to have a light network. The suggested structure is shown in Table III, this model only has one convolutional layer with 16 channels followed by a Max pooling layer in CAE part and the classifier part composes two convolutional layers with 32 and 64 channels.

*B. Impact of Region Gridding on localization performance*

In this sub-section, the effect of region gridding on localization performance is examined. The region gridding has a parameter $L$ which shows the length of each square in the localization area. The CAE-CNNLoc results for different amounts of $L$ are reported in Table IV in which the localization accuracy and model size are compared. Based on the given data, a model with $L = 7$ has the lowest error in location prediction and also the model parameters and size

Table III
Suggested network model structure

|  | layer | Layer feature | Output size |
|---|---|---|---|
| CAE | Input | - | 23x23x1 |
|  | Conv 2D | f:16, k:3 | 21x21x16 |
|  | Max pool | p: 2 | 7x7x16 |
| Classifier | Conv 2D | f:32, k:3 | 5x5x32 |
|  | Conv 2D | f:64, k:3 | 3x3x64 |
|  | Flatten | - | 576 |
|  | Dense | - | Num classes |

F: filter size, k: kernel size, p: pool size show a considerable reduction. It is reasonable because by increasing $L$ the number of classes is decreased which leads to a reduction in model parameters and consequently it reduces the model size. Additionally, longer $L$ generates a bigger square covering more points that their position is actually considered the unique coordinate of the square; hence it increases the localization error. Therefore, it reveals a trade-off between localization accuracy and model size, so based on the application we can adjust $L$ to achieve the best performance.

Table IV
Region Gridding effects on localization performance

| L | Euclidean Mean error | Building hitrate | Floor hitrate | Num Classes | File Size (MB) |
|---|---|---|---|---|---|
| 1 | 10.369 | 0.994 | 0.892 | 1,834 | 1.04 |
| 3 | 10.024 | 0.994 | 0.892 | 1,496 | 0.88 |
| 5 | 10.364 | 0.995 | 0.901 | 1,217 | 0.72 |
| 7 | 9.524 | 0.994 | 0.905 | 823 | 0.50 |
| 10 | 12.285 | 0.995 | 0.900 | 766 | 0.46 |
| 20 | 10.774 | 0.995 | 0.915 | 285 | 0.19 |
| 30 | 11.703 | 0.996 | 0.917 | 259 | 0.18 |
| 50 | 15.511 | 0.996 | 0.919 | 107 | 0.09 |

*C. Comparison with the existing methods*

Now, to evaluate the performance efficiency of CAE-CNNLoc method, we compare it with some recent studies. Some of the existing studies just focus on the building and floor accuracy, but in this study in addition to the building and floor accuracy, the positioning mean error has also been investigated. Table V reports the localization accuracy of CAE-CNNLoc and some related methods. It can be observed that the proposed method achieves accuracy in a range of other methods. In this regard, the floor hitrate of CAE-CNNLoc is 0.90, the positioning mean error is 9.52 and finally, like some of the other methods, building accuracy is approximately 100%.

*D. Noise resistance of CAE-CNNLoc*

WiFi signals in indoor environments are really vulnerable to noise; hence the RSSI value changes easily based on different conditions mentioned in sub-section III-B-1. Therefore, for indoor localization methods, it is critical to

Table V
CAE-CNNLoc in comparison with other methods

|  | Euclidean Mean error | Building hitrate | Floor hitrate | Model |
|---|---|---|---|---|
| **CAE-CNNLoc** | 9.52 | 0.994 | 0.905 | CAE+2D CNN |
| **CNNLoc** [13] | 11.78 | 1 | 0.96 | SAE+1D CNN |
| **Scalable DNN** [27] | 9.27 | 1 | 0.93 | SAE+DNN |
| **2D-CNN** [28] | - | 0.958* | | 2D CNN |
| **SAE+Classifier** [26] | - | 0.911* | | SAE+DNN |
| **Baseline** [25] | 7.9 | 0.899* | | KNN |
| **RTLS@UM** [34] | 6.2 | 1 | 0.94 | Filtering +KNN |

*The building and floor classes are combined

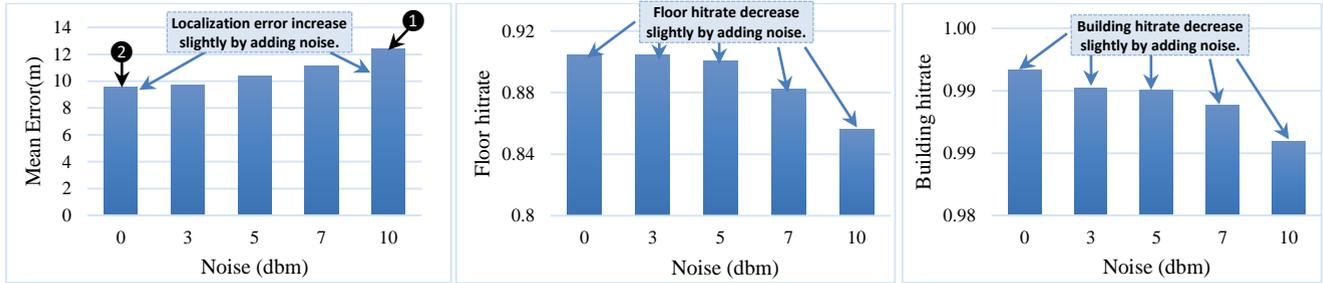

Figure 7: Effects of noise on CAE-CNNLoc performance.

resist noise. In this sub-section, we check out the performance of our model by adding noise to the data. For this aim, we randomly add 3, 5, 7 and 10 dBm noise to the test data and evaluate the model's accuracy. As is shown in Figure 7, the localization accuracy is declined by adding noise to signals which is reasonable; see ❶ and ❷, but it must be noted that in the worst case with 10 dBm noise, CAE-CNNLoc model shows 12.4 meter of error showing about 3 meters increase in error compared to the noise of 0 dBm. Therefore, these results properly verify the denoising feature of the convolution auto-encoder part of CAE-CNNLoc model.

*E. CAE-CNNLoc performance on android smartphone*

As mentioned before, the main purpose of this paper is to propose an on-device indoor localization model which can be run on the user's device. Thus, we examined the performance of CAE-CNNLoc model in the real world on an android smartphone. In this regard, the proposed CAE-CNNLoc model was tested on Redmi Note 8 and the results were reported. As two critical factors in on-device implementation, inference time (latency) and model size were taken into consideration and the impact of different techniques were reported. The Redmi Note 8 was powered by the Qualcomm Snapdragon 665 chipset, which featured an octa-core CPU clocked at up to 2.0 GHz. It came equipped with 4GB of RAM.

**Region gridding:** First, the effect of region gridding method was examined; in this regard, the results of the proposed method with different amounts of *L* are depicted in Figure 8. From sub-section IV-B, it is clear that by increasing *L*, the number of classes is decreased, leading to a remarkable reduction in network parameters which consequently declines the latency and network size as shown in Figure 8. For example, by set *L*=10, we can decrease the latency by about 2 times (50 µs) and reduce the network size over 2 times (460 KB) while the localization accuracy declines just nearly 2m compared with *L*=1. Hence, the region gridding method is a suitable way for on-device WiFi fingerprinting localization.

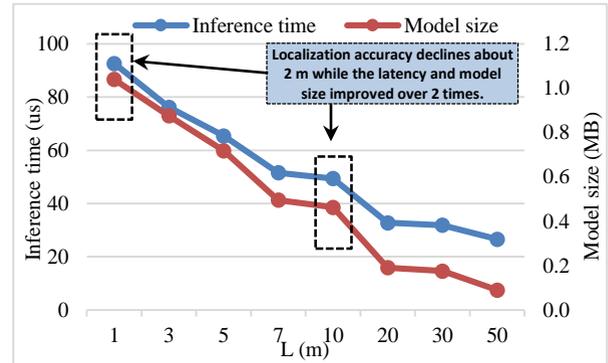

Figure 8: Effects of region gridding on localization inference time and model size.

**Quantization:** In this paper, we utilize float16 and int8 quantization which their results are shown in Figure 9. First, by using float16 quantization, weights type change from float32 to float16. As is clear from

Figure 9, for float16 quantization, although the network size reduces about 2 times, it does not affect localization accuracy and inference time. Moreover, Int8 quantization has the ability to decline the network size over 3.9 times and improve the latency by nearly 2 times without any effective reduction in localization accuracy.

*F. CAE-CNNLoc Scalability*

In this sub-section, CAE-CNNLoc model is applied on two other public WiFi fingerprinting datasets to evaluate the scalability of the proposed model. Table VI shows the

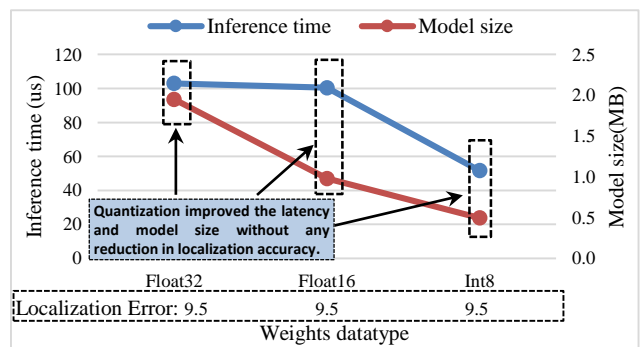

Figure 9: Effects of quantization on localization error, latency and model size.

Table VI
Scalability of CAE-CNNLoc model

| Dataset | Best of CAE-CNNLoc | | | Best of other DNN-based studies | | |
|---|---|---|---|---|---|---|
| | Mean error | Building hitrate | Floor hitrate | Mean error | Building hitrate | Floor hitrate |
| UJIIndoorLoc | 9.5 | 0.994 | 0.905 | 9.27 | 1 | 0.93 |
| Tampere | 10.24 | - | 0.889 | 10.88 | - | 0.94 |
| UTSIndoorLoc | 7.7 | - | 0.92 | 7.60 | - | 0.946 |

*Best of UJIIndoorLoc from [27], Tampere and UTSIndoorLoc from [13]

performance of CAE-CNNLoc on Tampere and UTSIndoorLoc datasets compared with recent studies. As is clear from this table, our model has a good performance and is close to the best of other DNN-based studies for the mentioned datasets, proving the scalability of CAE-CNNLoc.

## V. Discussion

As mentioned, in this study we transformed the localization problem from a regression to a classification task and benefited from region gridding to define the classes. It means that each cell (generated by the gridding technique) is considered as a class and all the points in that cell have the same coordination since the exact location is not really needed for most of the location-based services.

To this end, we applied region gridding on the training sets. However, our analysis revealed that there are lots of points in the test set that are not placed on the defined cell during the region gridding. This means that these points belong to some classes that are not seen by the model during training, and this is a reason that increases the localization error in our proposed classification-based method. This is depicted in Figure 10, as it is obvious there are lots of points from test set (red points) that are in separate cells without any points from train set to cover (blue/black points); meaning that these cells are not considered as a separate class for the model during training phase. Since in this study we transformed the localization from a regression to a classification task, we need to have a dataset which aligns with classification problem criteria. Thus, we have made some modifications to the dataset to better fit the requirements of the classification task. To this end, in the Dataset and Input Preparation phase (sub-section III-A), we first combine the training and testing sets; then the combined sets were randomly divided into training, validation and testing and then gridding method was used to define cells. By doing so, we can ensure that all the possible classes (cells) are seen by the model and have candidates during training, leading to a significant reduction in localization error as reported in Table VII. As is obvious, from the table, for UJIIndoorLoc the floor hitrate increased to 99%, the building hitrate is almost 100% and more importantly, the Euclidean error decreased to 2.36m.

Table VII
Suggested network model structure

| | Mean error | Building hitrate | Floor hitrate |
|---|---|---|---|
| UJIIndoorLoc | 2.36 | 0.9992 | 0.993 |
| Tampere | 6.30 | - | 0.961 |
| UTSIndoorLoc | 1.72 | - | 0.996 |
| SBUK-D | 1.73 | - | 0.985 |

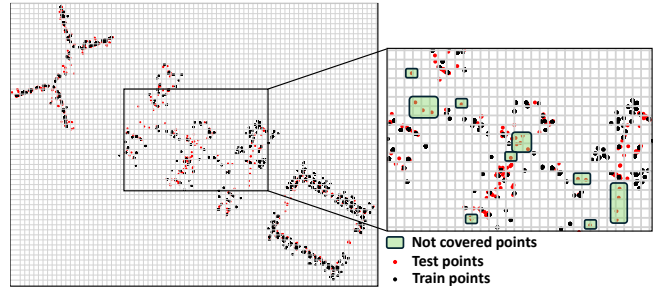

Figure 10: There are several isolated points (red points) in the test set that are not covered by the train set points (black points).

### A. Experiments on our dataset (SBUK-D dataset)

In this sub-section, we evaluate the performance of the proposed network model for localization in our department. For this aim, we generated the WiFi fingerprinting dataset named SBUK-D. The details of this dataset are as follows:

**SBUK-D dataset:** The authors gathered this dataset in the Engineering department at the Shahid Bahonar University of Kerman (SBUK). This building includes 3 floors that cover nearly 11,500 $m^2$ region. The position information consists of Floor Id, X and Y, and there are 198 different APs in the building used to record the RSSI value in each location. The recorded RSSI are between -100 dBm to 0 dBm, and like other datasets, inaccessible APs are set to 100. The dataset has 2292 samples for 70 different locations collected with 4 Android smartphones.

Table VII also reports the results of the CAE-CNNLoc on the SBUK-D dataset in comparison with three different datasets. As is clear from this table, CAE-CNNLoc achieved just 1.73 m error with over 98 percent accuracy in predicting the floor of SBUK-D dataset. Moreover, the model size for our dataset is about 103 KB and the inference time on an Android phone is 198 μs.

## VI. CONCLUSION

This paper proposed a light WiFi fingerprinting localization method named CAE-CNNLoc to estimate user position in indoor environments. CAE-CNNLoc is made-up the Convolutional Neural Network (CNN) network joined Convolutional Auto-encoder (CAE) that leads to a significant reduction in the input dimension and the model sensitivity to the input fluctuation. The proposed model can be easily run on edge devices e.g. smartphones, leading to a remarkable improvement in localization performance in terms of latency, and user' privacy. The experimental results illustrate that the proposed model outperforms other studies considering that it is an edge-based method. In this regard, for UJIIndoorLoc dataset, the model with int8 quantization is just 0.5 MB in size and obtains about 51 µs in inference time with 9.5m positioning error. Although localization performance is in the range of other studies, this is an edge-based system meaning that eliminates cloud/server disadvantages from the localization process. Besides, the proposed model shows significant performance on our new dataset named SBUK-D with 1.73 m positioning error, 98% accuracy of floor detection, inference time of 198 µs and the model size of 103 KB.

## ACKNOWLEDGMENT

This work has been supported in part by the NYUAD Center for Interacting Urban Networks (CITIES), funded by Tamkeen under the NYUAD Research Institute Award CG001, and Center for Artificial Intelligence and Robotics (CAIR), funded by Tamkeen under the NYUAD Research Institute Award CG010. Moreover, the first author acknowledges that this work was conducted while he was a researcher at Reliable & Smart Systems Laboratory, Shahid Bahonar University of Kerman, Iran.

## REFERENCES

[1] A. Basiri *et al.*, "Indoor location based services challenges, requirements and usability of current solutions," *Comput. Sci. Rev.*, vol. 24, pp. 1–12, May 2017, doi: 10.1016/j.cosrev.2017.03.002.

[2] J. Xiao, Z. Zhou, Y. Yi, and L. M. Ni, "A survey on wireless indoor localization from the device perspective," *ACM Computing Surveys*, vol. 49, no. 2. Association for Computing Machinery, pp. 1–31, Jun. 2016. doi: 10.1145/2933232.

[3] Q. Niu, M. Li, S. He, C. Gao, S. H. Gary Chan, and X. Luo, "Resource-efficient and automated image-based indoor localization," *ACM Trans. Sens. Networks*, vol. 15, no. 2, pp. 1–31, Feb. 2019, doi: 10.1145/3284555.

[4] M. Saadi, Z. Saeed, T. Ahmad, M. K. Saleem, and L. Wuttisittikulkij, "Visible light-based indoor localization using *k-means* clustering and linear regression," *Trans. Emerg. Telecommun. Technol.*, vol. 30, no. 2, p. e3480, Feb. 2019, doi: 10.1002/ett.3480.

[5] L. Ciabattoni *et al.*, "Real time indoor localization integrating a model based pedestrian dead reckoning on smartphone and BLE beacons," *J. Ambient Intell. Humaniz. Comput.*, vol. 10, no. 1, pp. 1–12, Jan. 2019, doi: 10.1007/s12652-017-0579-0.

[6] H. Zhang, K. Liu, F. Jin, L. Feng, V. Lee, and J. Ng, "A scalable indoor localization algorithm based on distance fitting and fingerprint mapping in Wi-Fi environments," *Neural Comput. Appl.*, vol. 32, no. 9, pp. 5131–5145, May 2020, doi: 10.1007/s00521-018-3961-8.

[7] J. J. Pérez-Solano, S. Ezpeleta, and J. M. Claver, "Indoor localization using time difference of arrival with UWB signals and unsynchronized devices," *Ad Hoc Networks*, vol. 99, p. 102067, Mar. 2020, doi: 10.1016/J.ADHOC.2019.102067.

[8] A. A. Nazari Shirehjini and S. Shirmohammadi, "Improving Accuracy and Robustness in HF-RFID-Based Indoor Positioning with Kalman Filtering and Tukey Smoothing," *IEEE Trans. Instrum. Meas.*, vol. 69, no. 11, pp. 9190–9202, Nov. 2020, doi: 10.1109/TIM.2020.2995281.

[9] A. Kargar-Barzi and A. Mahani, "Obstacle-resistant hybrid localisation algorithm," *IET Wirel. Sens. Syst.*, vol. 10, no. 5, pp. 242–252, Oct. 2020, doi: 10.1049/IET-WSS.2020.0052.

[10] S. He and S. H. G. Chan, "Wi-Fi fingerprint-based indoor positioning: Recent advances and comparisons," *IEEE Commun. Surv. Tutorials*, vol. 18, no. 1, pp. 466–490, Jan. 2016, doi: 10.1109/COMST.2015.2464084.

[11] S. Khandker, J. Torres-Sospedra, and T. Ristaniemi, "Analysis of Received Signal Strength Quantization in Fingerprinting Localization," *Sensors*, vol. 20, no. 11, p. 3203, Jun. 2020, doi: 10.3390/s20113203.

[12] F. Zafari, A. Gkelias, and K. K. Leung, "A Survey of Indoor Localization Systems and Technologies," *IEEE Commun. Surv. Tutorials*, vol. 21, no. 3, pp. 2568–2599, 2019, doi: 10.1109/COMST.2019.2911558.

[13] X. Song *et al.*, "A Novel Convolutional Neural Network Based Indoor Localization Framework with WiFi Fingerprinting," *IEEE Access*, vol. 7, pp. 110698–110709, 2019, doi: 10.1109/ACCESS.2019.2933921.

[14] D. J. Suroso, P. Cherntanomwong, and P. Sooraksa, "Fingerprint-based Indoor Localization via Deep Learning," *ACM Int. Conf. Proceeding Ser.*, pp. 146–152, Mar. 2023, doi: 10.1145/3592307.3592330.

[15] Q. Wang *et al.*, "DarLoc: Deep learning and data-feature augmentation based robust magnetic indoor localization," *Expert Syst. Appl.*, vol. 244, p. 122921, Jun. 2024, doi: 10.1016/J.ESWA.2023.122921.

[16] F. Alhomayani and M. H. Mahoor, "Deep learning methods for fingerprint-based indoor positioning: a review," *Journal of Location Based Services*, vol. 14, no. 3. Taylor and Francis Ltd., pp. 129–200, Jul. 2020. doi: 10.1080/17489725.2020.1817582.

[17] D. Ciuonzo and P. Salvo Rossi, "Distributed detection of a non-cooperative target via generalized locally-optimum approaches," *Inf. Fusion*, vol. 36, pp. 261–274, Jul. 2017, doi: 10.1016/j.inffus.2016.12.006.

[18] Y. Guo, K. Huang, N. Jiang, X. Guo, Y. Li, and G. Wang, "An exponential-rayleigh model for RSS-based device-free localization and tracking," *IEEE Trans. Mob. Comput.*, vol. 14, no. 3, pp. 484–494, Mar. 2015, doi: 10.1109/TMC.2014.2329007.

[19] L. Zhao, H. Huang, X. Li, S. Ding, H. Zhao, and Z. Han, "An accurate and robust approach of device-free localization with convolutional autoencoder," *IEEE Internet Things J.*, vol. 6, no. 3, pp. 5825–5840, Jun. 2019, doi: 10.1109/JIOT.2019.2907580.

[20] L. Ma, M. Liu, H. Wang, Y. Yang, N. Wang, and Y. Zhang, "WallSense: Device-Free Indoor Localization Using Wall-Mounted UHF RFID Tags," *Sensors*, vol. 19, no. 1, p. 68, Dec. 2018, doi: 10.3390/s19010068.

[21] H. Zhang, S. Y. Tan, and C. K. Seow, "TOA-Based indoor localization and tracking with inaccurate floor plan map via MRMSC-PHD filter," *IEEE Sens. J.*, vol. 19, no. 21, pp. 9869–9882, Nov. 2019, doi: 10.1109/JSEN.2019.2926433.

[22] F. Gu *et al.*, "Indoor localization improved by spatial context - A survey," *ACM Comput. Surv.*, vol. 52, no. 3, pp. 1–35, Jul. 2019, doi: 10.1145/3322241.

[23] X. Wang, L. Gao, S. Mao, and S. Pandey, "CSI-Based Fingerprinting for Indoor Localization: A Deep Learning Approach," in *IEEE Transactions on Vehicular Technology*, Institute of Electrical and Electronics Engineers Inc., Jan. 2017, pp. 763–776. doi: 10.1109/TVT.2016.2545523.

[24] B. Wang *et al.*, "A Novel Weighted KNN Algorithm Based on RSS Similarity and Position Distance for Wi-Fi Fingerprint Positioning," *IEEE Access*, vol. 8, pp. 30591–30602, 2020, doi: 10.1109/ACCESS.2020.2973212.

[25] J. Torres-Sospedra *et al.*, "UJIIndoorLoc: A new multi-building and multi-floor database for WLAN fingerprint-based indoor localization problems," in *IPIN 2014 - 2014 International Conference on Indoor Positioning and Indoor Navigation*, Institute of Electrical and Electronics Engineers Inc., 2014, pp. 261–270. doi: 10.1109/IPIN.2014.7275492.

[26] M. Nowicki and J. Wietrzykowski, "Low-effort place recognition with WiFi fingerprints using deep learning," in *Advances in Intelligent Systems and Computing*, Springer Verlag, 2017, pp. 575–584. doi: 10.1007/978-3-319-54042-9_57.

[27] K. S. Kim, S. Lee, and K. Huang, "A scalable deep neural network architecture for multi-building and multi-floor indoor Localization Based on Wi-Fi Fingerprinting," *arXiv*, vol. 3, no.

1. arXiv, pp. 1–17, Dec. 2017. doi: 10.1186/s41044-018-0031-2.
[28] J. W. Jang and S. N. Hong, "Indoor Localization with WiFi Fingerprinting Using Convolutional Neural Network," in *International Conference on Ubiquitous and Future Networks, ICUFN*, IEEE Computer Society, Aug. 2018, pp. 753–758. doi: 10.1109/ICUFN.2018.8436598.
[29] G. Zhang, P. Wang, H. Chen, and L. Zhang, "Wireless Indoor Localization Using Convolutional Neural Network and Gaussian Process Regression," *Sensors*, vol. 19, no. 11, p. 2508, May 2019, doi: 10.3390/s19112508.
[30] Z. Liu, B. Dai, X. Wan, and X. Li, "Hybrid Wireless Fingerprint Indoor Localization Method Based on a Convolutional Neural Network," *Sensors*, vol. 19, no. 20, p. 4597, Oct. 2019, doi: 10.3390/s19204597.
[31] S. Ioffe and C. Szegedy, "Batch Normalization: Accelerating Deep Network Training by Reducing Internal Covariate Shift," in *Proceedings of the 32nd International Conference on Machine Learning*, F. Bach and D. Blei, Eds., in Proceedings of Machine Learning Research, vol. 37. Lille, France: PMLR, 2015, pp. 448–456.
[32] S. Han, H. Mao, and W. J. Dally, "Deep compression: Compressing deep neural networks with pruning, trained quantization and Huffman coding," in *4th International Conference on Learning Representations, ICLR 2016 - Conference Track Proceedings*, International Conference on Learning Representations, ICLR, Oct. 2016.
[33] E. Lohan, J. Torres-Sospedra, H. Leppäkoski, P. Richter, Z. Peng, and J. Huerta, "Wi-Fi Crowdsourced Fingerprinting Dataset for Indoor Positioning," *Data*, vol. 2, no. 4, p. 32, Oct. 2017, doi: 10.3390/data2040032.
[34] A. Moreira, M. J. Nicolau, F. Meneses, and A. Costa, "Wi-Fi fingerprinting in the real world - RTLS@UM at the EvAAL competition," in *2015 International Conference on Indoor Positioning and Indoor Navigation, IPIN 2015*, Institute of Electrical and Electronics Engineers Inc., Dec. 2015. doi: 10.1109/IPIN.2015.7346967.

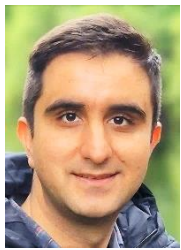

**Amin Kargar Barzi** is a Ph.D. student at Tyndall National Institute, University College Cork, Cork, Ireland. In the past, he worked as a researcher in the Reliable and Smart System (RSS) lab, Shahid Bahonar University of Kerman (SBUK), Iran. He received the B.Sc. degree in Electrical Engineering and the M.Sc. degree in Electronic Engineering both from SBUK. His research interests include machine learning, deep learning, tinyML, edge AI, machine vision and wireless smart systems.

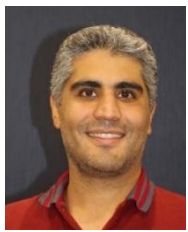

**Ebrahim Farahmand** (S'20) is an Associated Research at the Department of Electrical Engineering, Shahid Bahonar University of Kerman (SBUK), Iran. He received the B.Sc. degree in Electrical Engineering- Communication systems in 2012 and the M.Sc. degree in Electrical Engineering-Electronics in 2016, both from Shahid Bahonar University of Kerman (SBUK). From 2016 to present, he participated as a Research Associate in the Reliable and Smart System (RSS) lab, SBUK, Iran. His research interests include brain-inspired computing, deep learning, tinyML approximate computing, machine learning accelerator, Fault-tolerant design and Network systems.

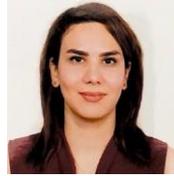

**Nooshin Taheri Chatrudi** is a Ph.D. student at the College of Health Solutions, Arizona State University (ASU). Currently, I am working under as a PhD student at the Embedded Machine Intelligence Lab (EMIL) of ASU. I received my M.Sc. degree in Communication Engineering with a focus on machine learning algorithms from the Shahid Bahonar University of Kerman (SBUK). My research interests include machine learning, clinical informatics, and health monitoring system development.

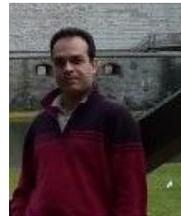

**Ali Mahani** received the B.Sc. degree in electronic engineering from Shahid Bahonar University of Kerman, Iran, in 2001, The M.Sc. and Ph.D. degrees both in Electronic Engineering from Iran University of Science and Technology (IUST), Tehran, Iran, in 2003 and 2009 respectively. Since then he has been with the electrical engineering department of Shahid Bahonar University of Kerman, where he is currently an associate professor. His research interests focus on Fault-tolerant design, FPGA-based accelerators, approximate digital circuits, stochastic computing and Networked System.

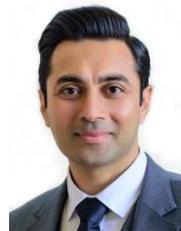

**Muhammad Shafique** (M'11 - SM'16) received the Ph.D. degree in computer science from the Karlsruhe Institute of Technology (KIT), Germany, in 2011. Afterwards, he established and led a highly recognized research group at KIT for several years as well as conducted impactful collaborative R&D activities across the globe. In Oct.2016, he joined the Institute of Computer Engineering at the Faculty of Informatics, Technische Universität Wien (TU Wien), Vienna, Austria as a Full Professor of Computer Architecture and Robust, Energy-Efficient Technologies. Since Sep.2020, Dr. Shafique is with the New York University (NYU), where he is currently a Full Professor and the director of eBrain Lab at the NYU-Abu Dhabi in UAE, and a Global Network Professor at the Tandon School of Engineering, NYU-New York City in USA. He is also a Co-PI/Investigator in multiple NYUAD Centers, including Center of Artificial Intelligence and Robotics (CAIR), Center of Cyber Security (CCS), Center for InTeractIng urban nEtworkS (CITIES), and Center for Quantum and Topological Systems (CQTS).

His research interests are in AI & machine learning hardware and system-level design, brain-inspired computing, quantum machine learning, cognitive autonomous systems, wearable healthcare, energy-efficient systems, robust computing, hardware security, emerging technologies, FPGAs, MPSoCs, and embedded systems. His research has a special focus on cross-layer analysis, modeling, design, and optimization of computing and memory systems. The researched technologies and tools are deployed in application use cases from Internet-of-Things (IoT), Smart Cyber-Physical Systems (CPS), and ICT for Development (ICT4D) domains. Dr. Shafique has given several Keynotes, Invited Talks, and Tutorials, as well as organized many special sessions at premier venues. He has served as the PC Chair, General Chair, Track Chair, and PC member for several prestigious IEEE/ACM conferences. Dr. Shafique holds one U.S. patent, and has (co-)authored 6 Books, 10+ Book Chapters, 300+ papers in premier journals and conferences, and 50+ archive articles. He received the 2015 ACM/SIGDA Outstanding New Faculty Award, the AI 2000 Chip Technology Most Influential Scholar Award in 2020 and 2022, the ASPIRE AARE Research Excellence Award in 2021, six gold medals, and several best paper awards and nominations at prestigious conferences. He is a senior member of the IEEE and IEEE Signal Processing Society (SPS), and a member of the ACM, SIGARCH, SIGDA, SIGBED, and HIPEAC.